\documentstyle[twocolumn,epsf,prl,aps]{revtex}
\begin{document}
\draft

\title{
Dynamics of nanojets
    }

\author{Jens Eggers}
\address{
Universit\"{a}t Gesamthochschule Essen, Fachbereich Physik,
45117 Essen, Germany  }
\maketitle

\begin{abstract}
We study the breakup of a liquid jet a few nanometers 
in diameter, based on a stochastic differential equation
derived recently by Moseler and Landman [Science {\bf 289},
1165 (2000)]. In agreement with their simulations, we
confirm that noise qualitatively changes the characteristics
of breakup, leading to symmetric profiles. Using the path
integral description, we find a self-similar profile that
describes the most probable breakup mode. As illustrated
by a simple physical argument, noise is the driving force
behind pinching, speeding up the breakup to make surface
tension irrelevant. 

\end{abstract}
Given the current trend for miniaturization, it is natural 
to ask about the behavior of liquids on increasingly small scales,
where thermal fluctuations eventually become important. 
In the case of free surfaces, relevant for all printing
and microstructuring applications \cite{W01}, fluctuations 
can expected to be even larger than in confined geometries. 
Their importance 
is estimated by comparing the thermal energy $k_B T$ with the
surface tension $\gamma$. Thus for structures whose size is in the order of 
the thermal length scale $\ell_T = (k_B T/\gamma)^{1/2}$
\cite{SBN}, usually about a nm at 
room temperature, fluctuations should make a leading-order 
contribution. Predicting the (typical) behavior of fluids on
the nanoscale is therefore an extremely challenging statistical physics 
problem, since the noise can no longer be treated in a standard
perturbative fashion \cite{Ris}. Instead, non-perturbative methods
are called for to properly account for thermal fluctuations. 

In the absence of experiments that could directly measure fluid
flow on the nanoscale, molecular dynamics (MD) \cite{AW70,KBW}
computations are an ideal tool. Previous analyses of
drops and threads \cite{GH91,KB93,K98} convincingly showed 
hydrodynamic behavior on the nanometer scale, and agreement 
with breakup times and growth rates as predicted by linear 
theory \cite{E97}. Recently \cite{ML} a realistic molecular 
dynamics simulation of a jet of propane issuing from 
a nozzle 6 nm in diameter  was performed, which also payed 
close attention to the nonlinear dynamics close to breakup. 
Remarkably, a coherent jet develops, from which drops separate at
slightly irregular intervals to form a quasi-stationary decaying
state. If nozzles of this size were to be built, this 
opens dazzling perspectives of transporting current printing and
structuring techniques into the nanoworld. 

However, a careful analysis of the breakup process revealed \cite{ML} that 
the presence of noise qualitatively alters the character of 
the breakup. While the deterministic motion at a corresponding
Reynolds number forms elongated necks between 
successive drops \cite{E97}, noise leads to profiles 
symmetric around the pinch point. Thus satellite drops 
rarely form and quite surprisingly the distribution of drop 
sizes becomes narrower. 
In addition, on the nanoscale the motion of the minimum neck radius 
accelerates as breakup is approached, while the corresponding time 
dependence $ h_{min} = 0.03 (\gamma\rho/\nu)(t_0-t)$ is linear 
for deterministic pinching \cite{E93}. Here $\nu$ is the kinematic 
viscosity and $\rho$ the density of the fluid. Thus the theoretical challenge 
is to understand this qualitative change of behavior in a regime where noise
makes the leading contribution. 

To deal with the above set of problems, we use the continuum description 
given by \cite{ML}, which consists in adding a stochastic term 
to the lubrication description of a deterministic jet \cite{ED}.
The amplitude of the noise is fixed by the condition of detailed 
balance \cite{LL}. Detailed numerical simulations of the 
stochastic equation gave very convincing agreement with MD simulations. 
This means that hydrodynamics, at least when suitably generalized 
to include fluctuations, is fully capable 
of describing free-surface flows down to a scale of nanometers.

The coupled set of differential equations for the radius of the 
fluid jet $h(z,t)$ and the axial velocity $v(z,t)$, 
as derived in \cite{ML}, reads 
\begin{equation}
\label{ml1}
\partial_t h^2 + (h^2v)' = 0
\end{equation}
\begin{equation}
\label{ml2}
\partial_t (h^2v) + (h^2v^2)' = -G' + 
3 (h^2v')' + D(h\xi)' ,
\end{equation}
where the prime refers to differentiation with respect to the 
spatial variable. The first equation (\ref{ml1}) expresses mass
conservation, (\ref{ml2}) is the momentum balance.
All quantities are written in units of the 
intrinsic length and time scales
$\ell_{\nu}=\nu^2\rho/\gamma$, $t_{\nu}=\nu^3\rho^2/\gamma^2$,
respectively. For later convenience the Laplace pressure term 
is written in the form $G' = h^2\kappa'$, 
$\kappa$ being the mean curvature of the interface. 
The Gaussian Langevin force is uncorrelated in space and time, 
\begin{equation}
\label{white}
<\xi(z_1,t_1)\xi(z_2,t_2)> = \delta(z_1-z_2)\delta(t_1-t_2),
\end{equation}
and the dimensionless constant $D = (6/\pi)\ell_T/\ell_{\nu}$ measures 
the strength of the noise contribution. 

Since the derivative of the noise is an extremely singular 
quantity, it is useful to integrate (\ref{ml2}) once, setting 
$P = \int_0^z p(x) dx$,
where $p = h^2v$ is the momentum. Thus we arrive at the 
conserved form of the equations 
\begin{equation}
\label{cml1}
\partial_t h^2 = -P'' 
\end{equation}
\begin{equation}
\label{cml2}
\partial_t P = -(P^2/h^2)'  -G' + 
3 h^2(P'/h^2)' + Dh\xi .
\end{equation}

\begin{figure}
\caption{
The computed interface profile of a collapsing bridge of liquid 
propane at 185 K. All lengths are nondimensionalized by the initial 
bridge radius of 3 nm. The time interval between the profiles is 
150 ps. 
}
\begin{center}
    \leavevmode
    \epsfsize=0.4 \textwidth
    \epsffile{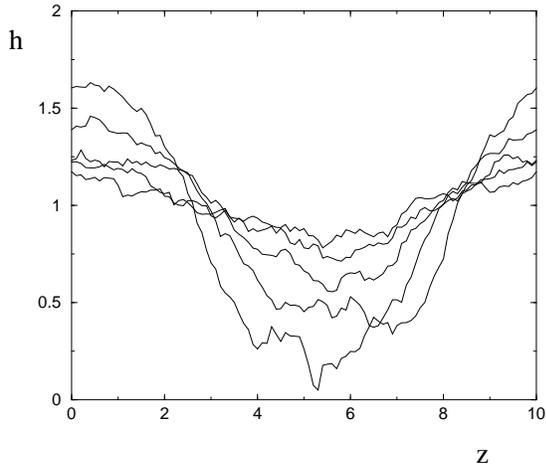}
  \end{center}
\label{prf}
\end{figure}

Figure \ref{prf} shows the collapse of a liquid thread of propane
6nm in diameter as given by (\ref{cml1},\ref{cml2}) 
with periodic boundary conditions. 
The results agree well with the computation shown in the 
supplementary Fig.1 of \cite{ML}, and 
in particular the profile remains close to being symmetric. 
Remembering that $P$ is the velocity of the fluid times a typical 
volume, the multiplicative factor in front of the random force 
in (\ref{cml2}) corresponds to a relative {\it increase} in noise
strength as pinch-off is approached. This provides us with a simple 
physical picture for an effective force generated by fluctuations. 
Namely, a random fluctuation which increases the thread radius 
also increases its effective mass, slowing done the motion. Any fluctuation
towards a smaller neck radius, on the other hand, accelerates the motion. 
On average, the fluctuations thus drive the thread towards breakup,
in fact more effectively than surface tension as we will see below.
One should also note the similarity of (\ref{cml2}) with the 
equation describing directed percolation \cite{JKO99}.

It is clear that conventional perturbative expansions around
the deterministic solution are doomed to fail in describing this 
mechanism, since the noise makes a dominant contribution. 
In fact, any average taken at a fixed 
time no longer makes sense for this problem, because there is a finite 
probability for pinch-off to have occurred, so the original formulation
ceases to be valid. Thus a valid description has to be conditioned 
on the event of breakup to take place at a fixed time $t_0$. 
It is then natural to ask for the {\it most probable} sequence 
of profiles that brings one from the initial condition to some 
final profile at time $t_0$, corresponding to a ``typical''
breakup event. From a standard path integral description, 
using the Gaussian properties of $\xi$, one finds \cite{Ris} that the 
probability for such a path is 
\begin{equation}
\label{W}
W\{h(z,t),P(z,t)\} \sim exp\{-\int_0^{t_0} dt L\} ,
\end{equation}
where the ``Lagrangian'' is
\begin{equation}
\label{L}
L = \frac{1}{2D^2} \int dz 
\frac{(\dot{P} + (P^2/h^2)' + G' - 3 h^2(P'/h^2)')^2}{h^2}.
\end{equation}

The first equation (\ref{cml1}) has no noise
in it and has to be treated as a constraint. To find
the most probable path {\it with fixed end-points} one derives 
the Euler-Lagrange equation \cite{Gra}
for the variation in $a=h^2$ and $P$, with the constraint
treated by adding a Lagrange multiplier 
$\tilde{a}(\dot{a} + P'')$ to L. It is somewhat more convenient
to pass to a ``Hamiltonian'' description, introducing
$\tilde{P} = \partial L/\dot{P}$ as the conjugate field, 
in the literature on critical phenomena also known as the 
``response'' field \cite{MSR,J76}.  From this one directly
finds the Hamiltonian equations 
\begin{eqnarray}\label{instanton}
&& \partial_t a = -P'' \nonumber \\
&& \partial_t P = -P'^2/a + 3a(P'/a)' + D^2a\tilde{P}\\
&& \partial_t \tilde{a} = -\frac{D^2}{2}\tilde{P}^2
- \tilde{P}P'^2/a^2 - 3(\tilde{P}P')'/a  \nonumber \\
&& \partial_t \tilde{P} = -2(\tilde{P}P'/a)'
 - 3((\tilde{P}a)'/a)' + \tilde{a}'' . \nonumber
\end{eqnarray}

The contribution $G$ from surface tension was dropped,
since it will turn out below that it is sub-dominant
near the pinch point. Introducing the transformation 
$\bar{P}=D^2\tilde{P}$ and $\bar{a}=D^2\tilde{a}$
the amplitude $D$ can be scaled out of the problem. 

``Optimal'' paths such as those described by (\ref{instanton}) 
have recently enjoyed some popularity in the statistical mechanics
literature \cite{FKLM,FB}. 
However, there are two conceptual difficulties 
associated with (\ref{instanton}). The first is that 
the equation for $\tilde{P}$ contains a term that corresponds
to {\it negative} diffusion, so it cannot be integrated forward 
in time. In \cite{FB} an ingeneous yet extremely elaborate 
computational scheme was developed 
to deal with this problem, based on an initial guess of
the {\it complete evolution} of the profiles. In subsequent
iterations, the physical fields were always integrated forward
in time, the conjugate fields backward in time. A second, 
perhaps more serious problem is that one does not know a priori
what the final profile is supposed to be, so it has to be
guessed. Once a solution is found, the probability of finding
a given final profile can be estimated by evaluating the 
probability of the total path. This evidently amounts 
to a daunting mathematical problem for a complicated 
system like (\ref{instanton}). 

Both problems can simultaneously be dealt with by assuming that 
the solution is {\it self-similar}, as found for the deterministic 
solution \cite{E93,RRR}. This means we assume solutions 
for small $|t'|=|t_0-t|$ to be of the form
\begin{eqnarray}\label{sim}
&& a(z,t) = |t'|^{2\alpha} \phi^2(\xi),\quad
P(z,t) = |t'|^{2\alpha} \psi(\xi) \nonumber   \\
&& \bar{a}(z,t) = |t'|^{-1} \Gamma(\xi), \quad
\bar{P}(z,t) = |t'|^{-1} \chi(\xi), \\
&& \xi = z/|t'|^{1/2}, \nonumber
\end{eqnarray}
where the exponent $\alpha$ remains to be determined. 
Plugging (\ref{sim}) into (\ref{instanton}), one finds
the similarity equations 
\begin{eqnarray}\label{simequ}
&& -2\alpha\phi^2 + \frac{\xi}{2}(\phi^2)' = -\psi'' \nonumber \\
&& -2\alpha\psi + \frac{\xi}{2}\psi' = -\psi'^2/\phi^2 + 
3\phi^2(\psi'/\phi^2)' + \phi^2\chi \\
&& \Gamma + \frac{\xi}{2}\Gamma' = -\frac{1}{2}\chi^2
- \chi\psi'^2/\phi^4 - 3(\chi\psi')'/\phi^2  \nonumber \\
&& \chi + \frac{\xi}{2}\chi' = -2(\chi\psi'/\phi^2)'
 - 3((\chi\phi^2)'/\phi^2)' + \Gamma'', \nonumber
\end{eqnarray}
where now the prime refers to differentiation with respect
to $\xi$. One notices immediately that (\ref{simequ}) is
invariant under the transformation $\phi\rightarrow A\phi$
and $\psi\rightarrow A^2\psi$, so in the following it is  
enough to look for solutions with $\phi(0)=1$.

Physically meaningful solutions of the set (\ref{simequ})
must match onto an outer solution that is slowly 
varying on the local timescale set by $|t'|$ \cite{E93}, which means
that $|t'|$ must drop out of the similarity description
for large arguments $\xi\rightarrow\pm\infty$. For the profile $\phi$ this 
means that it must grow like $\phi \sim \xi^{2\alpha}$ at infinity.
We are looking for {\it symmetric} solutions of (\ref{simequ}),
which were found not to exist for its deterministic counterpart \cite{E93}.
Amazingly, we found that the subset of solutions of (\ref{simequ}) with the 
correct asymptotic behavior obey the second order {\it linear}
equation
\begin{equation}
\label{lin}
\psi'' = \frac{2\alpha}{3}\psi - \frac{\xi}{6}\psi',
\end{equation}
and the other functions can be written in terms of 
$\psi$ as follows:
\begin{equation}\label{simf}
\phi^2 = \psi/3,\quad\Gamma = -27 (\psi'/\psi)^2,\quad
\chi = -18 (\psi'/\psi)' ,
\end{equation}
which is verified by substitution. Symmetric solutions of 
(\ref{lin}) with $\psi(0)=3$ are also related to the confluent hypergeometric 
function \cite{LLQu} by $\psi = 3 F(-2\alpha,1/2,-\xi^2/12)$.
\begin{figure}
\caption{
The symmetric similarity profile $\phi$ as given by a 
solution of (\ref{lin}),(\ref{simf}) with $\alpha=0.418$.
}
\begin{center}
    \leavevmode
    \epsfsize=0.4 \textwidth
    \epsffile{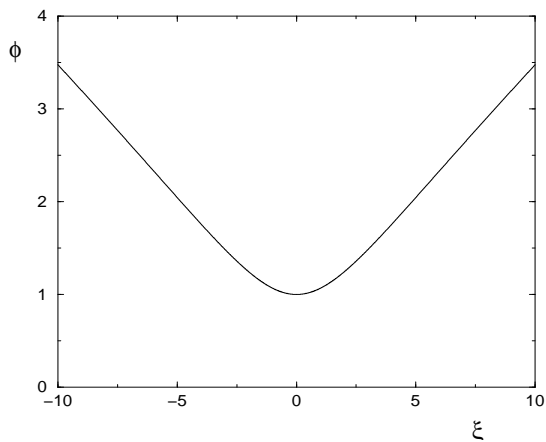}
  \end{center}
\label{phi}
\end{figure}

This gives a complete description of the physical solutions 
of (\ref{simequ}), but one still has to find the value of the 
exponent $\alpha$. To that end we compare the weight
(\ref{W}) of solutions with different $\alpha$, to find the one
for which $L$ is minimum. From (\ref{instanton})
we find that 
\[
L = \frac{1}{2D^2} \int_{-\ell}^{\ell} dz a \bar{P} + L_0,
\]
where the integration is over the small spatial region where 
the similarity description applies and $L_0$ is essentially
a constant as $|t'|\rightarrow 0$. 

Keeping the initial height $h_0$ of the liquid bridge constant, 
we find that the integral $\int_0^{t_0} L$ is dominated by contributions 
from $|t'|\rightarrow0$ if $\alpha<1/2$. Thus in similarity variables 
we have up to constants 
\begin{equation}
\label{opt}
\int_0^{t_0} dt L = \frac{h_0^2}{2D^2t_0^{1/2}}\frac{1}{2\alpha-1/2}
\int_{-\infty}^{\infty} d\xi \phi^2 \chi^2 .
\end{equation}
The remaining task is to minimize (\ref{opt})
as a function of $\alpha$. The decay of the argument of the 
integral is like $\xi^{4\alpha-4}$, so the integral converges
for $\alpha<3/4$. We conclude that (\ref{opt}) must have a 
minimum somewhere between $\alpha=1/4$ and 3/4.

To find it, one has to do the integral numerically for 
general $\alpha$ and finds 
$ \alpha_{min} \approx 0.418 < 1/2$, consistent with the assumptions made
above. In particular, $\alpha$ is smaller than 1, so surface
tension becomes asymptotically sub-dominant, and hence the singularity is
driven by noise alone. A more quantitative comparison with numerical 
simulations requires considerably better statistics than we are
presently able to accumulate. 

We have thus shown that the optimal path method can 
be used to reveal qualitatively new features of nanoscale flows, 
namely a speed-up in pinching and symmetric profiles. 
It seems that this method is particularly well suited to treat problems
arising in the new science of nanoscale devices. Typically one is 
interested in the most probable or 'typical' behavior that brings 
one to a certain end, say to have a device complete a given motion. 
I claim that the present method is tailored to this situation. 

\acknowledgements
I am indebted to Michael Brenner, who first suggested the use of
path integrals, and to Robert Graham and Walter Strunz for greatly 
clarifying my views on extremal solutions. I also owe very interesting
and helpful conversations to Mary Pugh and to Joachim Krug.

\end{document}